\magnification=\magstep1
\hsize=15 true cm
\vsize=22 true cm
\bigskip
\centerline{Evolution of structure functions in a chiral potential model}
\bigskip
\centerline{Kakali Roy-Maity}\centerline{and}
\centerline{Padmanabha Dasgupta}
\centerline{\it Department of Physics.}
\centerline{\it University of Kalyani.}
\centerline{{\it Kalyani,West Bengal.}741235}
\centerline{\it India.}
\centerline{E-mail:dasgupta@klyuniv.ernet.in}
\bigskip
\bigskip
\bigskip
\bigskip
\centerline{Abstract}
\noindent Evolution of structure functions is studied in a chiral potential model which
incorporates the pion as the chiral symmetry restoring field. Evolution equations
for the quark and pion densities are derived in the lowest order at large $Q^2$.
The splitting function for quark emission from point-like pion is flavor-dependent.
This is shown to lead to nontrivial evolution of the nonsinglet moments in the
next order which is consistent with the observed departure from the Gottfried
sum rule.
\bigskip
\bigskip
\bigskip
\bigskip
\leftline{PACS:12.40Qq,13.60Hb,14.20Dh }
\vfill\eject
\leftline{\bf 1\quad{ Introduction.}}\medskip
\par The observed departure from the Gottfried sum rule is difficult to explain
within the framework of perturbative QCD [1]. The effect indicates a flavor-asymmetry
in the sea quark distribution inside a nucleon which cannot be generated by the
Altarelli-Parisi evolution of the structure functions. The simplest alternative
is to bring in pions into the picture [2,3,4] and replace gluon emission from a
quark by pion emission [5]. Just having pions inside the nucleons cannot, of course,
explain the observed effect because the direct contribution of the pion density
drops out from the equation of evolution of the nonsinglet moment which is the
object of interest in the context of the Gottfried sum. A rather complete treatment of the
problem has been given by Ball and Forte [6] who have computed the non-perturbative
cross sections for the emission of pseudoscalar bound states from quarks and
obtained a satisfactory explanation of the observed phenomenon.
 Their work shows that the effect is essentially nonperturbative.
 Also, there is good theoretical ground to expect such evolution, even if
 the analysis of the experimental data is clouded with uncertainties.\par In this paper
we are interested in seeing whether such an evolution can result in the phenomenological
models of the
nucleons which have been devised to study the static properties of the nucleon
bound state. The question is nontrivial since the connection between these models
and the QCD-evolved structure functions is not at all obvious.On the other hand
,the models of the nucleon bound state are supposed to be phenomenological
manifestations of the nonperturbative QCD at the hadronic scale[16]. It is therefore
natural to try to see to what extent the models serve as effective theories and whether
they can be put to any use other than calculation of static properties of the
nucleon. As an example, we have chosen in this paper a chiral potential model incorporating point-like pions
and have calculated the relevant cross sections and the splitting functions using
the quark-pion coupling of that model. The choice of this particular model is
not due to its having any greater intrinsic merit than other existing models,
but due to the simplcity of its quark-pion hamiltonian.\par We report here only the low
order perturbation results at large $Q^{2}$ and show that a nontrivial evolution
of the nonsinglet distribution can indeed result from this model.

\leftline{\bf 2\quad{\bf The model.}}\medskip
\par We consider, with some  modification, a phenomenological independent-quark model of baryons
studied by Jena and Panda[6,7] (see also [8-15]) who could obtain the static electromagnetic
properties of the nucleon, the axial vector coupling constant in beta decay
and the pion-nucleon coupling constant in resonable agreement with the experiments.
The quarks in this model move independently in a confining potential $V_q(\vec r)$
which is an equal mixture of a vector and a scalar part:
$$V_q(\vec r)={1\over 2}(1+\gamma^o)V(\vec r)\eqno(2.1)$$
with $$V(\vec r)=a^{2}r+V_\circ,~a>o$$
The non-invariance of the quark mass term and the scalar part of $V_q(\vec r)$
under SU(2) chiral transformation is sought to be adjusted by introducing
a pion  field interacting with the quarks through a linearised term in the
lagrangian density
$${\cal L}_{I}^{\pi}(x)={i\over f_\pi}G(\vec r)\bar {\psi_{q}}(x)\gamma^5{\bf{\tau.\phi(x)}}\psi_{q}(x)\eqno (2.2)$$
$$G(\vec r)=m_q+{V(\vec r)\over 2 }\eqno(2.3)$$
$m_{q}$ is the quark mass and $f_{\pi}$ is the pion decay constant.The quark wavefunction of the baryonic core
is determined from$V_{q}(\vec r)$;in the ground state configuration$$\psi_{q}(\vec r)=N_{q}\pmatrix{\phi_{q}(r) & &
\cr -i\sigma.\hat r\over \lambda_{q} & d \phi_{q}(r)\over dr}\chi^{\uparrow}\eqno(2.4)$$ $\phi_q(r)$ turns
out to be expressible in terms
of the Airy function.\par The static properties of the baryons calculated from$V_{q}(\vec r)$ alone
(with $a\simeq 0.343$ Gev,$V_\circ\simeq -0.506$ Gev)are
alredy close to the observed values so that, on a phenomenological level, the quark-pion coupling is
 expected to be small making perturbative calculations meaningful. On the other hand,this is an example of a potential model
which allows one to calculate pion emission and absorption by quarks and their effect on the structure functions.
We show this by considering, as the custom is, a quark beam which is being probed
by a virtual photon, obtaining the splitting functions from the lowest-order cross sections.\par
The relevant cross sections are (a) $\sigma(\gamma^*q\rightarrow {q^\prime}\pi)$,
~(b) $\sigma(\gamma^{*}q\rightarrow{\pi}q^{\prime})$and (c) $\sigma(\gamma^*\pi
\rightarrow q\bar{q})$.
\par The processes (a) and (b) contain emission of quarks from quarks
and pions,and lead to the splitting functions$P_{qq}$ and$P_{q\pi}$.The process(c)
gives the splitting function$P_{{\pi}q}$.
\par The external potential G having a long range part in the original model
of Panda and Jena necessitates the introduction of an infrared cutoff parameter
$\lambda$ in the three momentum of the static gluon.Physically, this is fixed
by the size of the baryons.With this modification, the effective $qq\pi$
vertex factor for perturbation calculations in this model becomes
$$\Gamma_{qq\pi}(p,p^\prime)={1\over 2f_\pi}\left[V_\circ-{4a^2(\vert\vec p-
\vec p^\prime \vert -\lambda)\over \pi\lambda\vert\vec p-\vec p^\prime\vert}\right]
\gamma^5\eqno(2.5)$$ where $p$ and $p^\prime$ are the quark four-momenta going
into and out of the vertex.  \vfill\eject
\leftline{\bf 3 \quad Virtual photon cross sections in the large $Q^2$ limit.}
\medskip \par The
evolution of quark and pion densities in this model can be obtained by first
calculating the virtual photon cross sections listed in the previous section.
The lowest order perturbation calculation using the vertex factor(2.5) is
straightforward in the small transverse momentum and large $Q^2$ limit neglecting
the quark and pion masses.\par In the following  ~$p$, $p^\prime$,$k$,$q$ denote
the (center-of-mass)four-momenta of the quarks $q$, $q^\prime$, the pion and the
photon.For the process(a),we define the invariant variables  $s_a=(q+p)^2$,  $t_a=(p^\prime-q)^2,
$ ~$u_a=(p-p^\prime)^2$. Then in the leading order, the cross
section $\sigma(\gamma^*q\rightarrow q^\prime\pi)$ is
$${d\sigma\over d\Omega^\prime}={2\pi\alpha\over s_a^2}e_{q^\prime}^2\left({-s_a
\over t_a}-{2(s_a+Q^2)Q^2\over s_at_a}\right)\left[V_\circ-{4a^2\over \pi\lambda}\right]^2\eqno(3.1)$$
\par Defining $s_b=(q+p)^2$,$t_b=(q-k)^2$ and $u_b=(p^\prime -q)^2$, the cross
section for the process $\gamma^*q\rightarrow \pi q^\prime$ is
$${d\sigma\over d\Omega_\pi}={2\pi\alpha\over s_b^2}{e_{q^\prime}^2\over 4f_\pi^2}
\left[V_\circ-{4a^2\over {\pi\lambda}}\right]^2{2Q^2\over -t_b}\eqno(3.2)$$
\par Defining $s_c=(q+k)^2$~,~$t_c=(q-p^\prime)^2$~and~$u_c=(p^\prime-k)^2$,
we find, to leading order in $Q^2$, the cross section for the process
$\gamma^{*}\pi\rightarrow  q^\prime \bar{q}$~
$${d\sigma\over d\Omega^\prime}={2\pi\alpha\over s_c^2}{e_{q^\prime}^2\over f_\pi^2}
\left[V_\circ-{4a^2\over{\pi\lambda}}\right]^2{1\over 2}\left[{\left(u_c\over t_c\right)}+
{2s_cQ^2\over t_c{u_c}}{{\left(e_\pi\over e_{q^\prime}\right)^2-1}}\right]\eqno(3.3)$$
To facilitate the extraction of the splitting functions, we express these cross
sections in terms of the kinematic variables $z$ and $p_T$ for each process which
are related to the  momentum variables by$$z\simeq {Q^2\over s+Q^2}\eqno(3.4)$$
  $$p_{T}^2\simeq {-st\over s+Q^2}\eqno(3.5)$$
the relations being true for small  $t$ and large $Q^2$. Under these approximations,
$${d\sigma(\gamma^{*}q\rightarrow q^{\prime}\pi)\over dp_T^2}=
e_{q^\prime}^2\sigma_\circ{\alpha_{\pi}\over p_T^2}\left[{1+z^2}\over {1-z}\right]
\eqno(3.6)$$
$${d\sigma(\gamma^*q\rightarrow \pi q^{\prime})\over dp_T^2}
=e_{\pi}^2\sigma_{\circ}{\alpha_{\pi}\over{p_T^2}}{2z}\eqno(3.7)$$
$${d\sigma(\gamma^*\pi\rightarrow q^{\prime}\bar{q})\over dp_T^2}
=e_{q^\prime}^2\sigma_\circ{\alpha_{\pi}\over{p_T^2}}{1\over 2}\left[(z^2+(1-z)^2)+
\left(e_{\pi}^2\over e_{q^\prime}^2\right)z(1-z)\right]\eqno(3.8)$$
where $\sigma_\circ=4\pi^2\alpha /s$ and  $$\alpha_{\pi}={1\over {2\pi f_\pi^2}}\left[V_\circ-
{4a^2\over \pi\lambda}\right]^2\eqno(3.9)$$ which is an effective coupling 
constant of the model in the limit of large $Q^2$.
\medskip\leftline{\bf 4.\quad {\bf The evolution equations and splitting functions.}}
\medskip
Integrating (3.6)-(3.8) with respect to $p_T^2$ (with an infra-red cutoff),
we can now easily get the evolution of the structure function $F_2(x,Q^2)$.
The quark densities are then found to satisfy the following equations.
$${du(x)\over dlogQ^2}=\alpha_{\pi} \int_{x}^{1}{dy\over y}\left[u(y)P_{uu}(z)+
d(y)P_{ud}(z)+\pi^\circ(y)P_{ u\pi^\circ}(z)
+\pi^+(y)P_{u\pi^+}(z)\right]\eqno(4.1)$$
$${d{\bar u}(x)\over dlogQ^2}=\alpha_{\pi}\int_{x}^{1}{dy\over y}\left[\bar {u}(y)P_{\bar u
\bar u}(z)+\bar{d}(y)P_{\bar u\bar d}(z)
+\pi^\circ (y) P_{\bar u\pi^\circ}(z)+\pi^-(y)P_{\bar u \pi^-}(z)\right]\eqno(4.2)$$
and two similar equations for  $d(x,Q^2)$ and $\bar{d}(x,Q^2)$. $P_{ab}$ stands
for the splitting function for $b\rightarrow a $ and $z=x/y$.
The pion densities are found to satisfy
$$ {d\pi^+(x)\over dlogQ^2}=\alpha_{\pi}\int_{\circ}^{1}{dy\over y}\left[u(y)P_{\pi^+u}(z)
+\bar{d}(y)P_{\pi^+\bar d}(z)\right]\eqno(4.3)$$
$${d\pi^-(x)\over dlogQ^2}=\alpha_{\pi}\int_{\circ}^{1}{dy\over y}\left[d(y)P_{{\pi^-}d}
(z)+\bar{u}(y)P_{{\pi^-}u}(z)\right]\eqno(4.4)$$
$$ {d\pi^\circ(x)\over dlogQ^2} =\alpha_{\pi}\int_{\circ}^{1}{dy\over y}\left[u(y)P_{
{\pi^\circ}u}(z) +d(y)P_{{\pi^\circ}d}(z)+\bar{u}(y)P_{{\pi^\circ}\bar u }(z)+
\bar{d}(y)P_{{\pi^\circ}\bar d}(z)\right]\eqno(4.5)$$
The splitting functions are given by
$$P_{{q^\prime}q}(z)={1\over 2}{1+z^2\over 1-z},~~  P_{{\pi}q}(z)=2z$$
$$P_{{q^\prime}\pi}={1\over 2}\left[(z^2+(1-z)^2)+\left(e_\pi\over e_q^\prime
\right)^2z(1-z)\right]\eqno(4.6)$$
The splitting functions obviously satisfy the requirement of charge conjugation
invariance : $P_{qq}=P_{\bar q\bar q},~ P_{q\pi^+}=P_{\bar q\pi^-},~ P_{q\pi^\circ}
=P_{\bar q\pi^\circ}$ etc.As given in (4.6), they do not satisfy the fermion
number conservation, but this could be ensured by the usual regularization procedure.
\medskip
\leftline{\bf { 5.Evolution of the nonsinglet distribution.}}\medskip
\par The first thing one notes about the results given above is that even in
the lowest order the evolution of individual quark densities is flavor-dependent.
However, this alone can not explain the observed evolution of the Gottfried
sum. Firstly, the first order calculation of the cross sections will always lead
to $P_{{\bar q}q}=P_{q\bar q}=0$.The contribution of $P_{{q^\prime}\pi}$
drops out from the evolution of the nonsinglet distribution, and hence the
dependence of $P_{q\pi}$ does not matter at all in this context.
\par Secondly, the drastic large $Q^2$ approximation used to obtain the results of the previous
section will not lead to any evolution of the anomalous dimensions obtained by taking
moments of the splitting functions.Hence, the Gottfried sum remains unaffected
as it is controlled by the anomalous dimension for the first moment. \par Therefore,
in this model it is essential to consider both higher order perturbation terms and
low $Q^2$ corrections. The splitting function $P_{\bar{q}q}$ is generated by the
second order process and is seen to be of the form
$$P_{\bar{q}q}(z)\simeq {\alpha_{\pi}}\left[{1\over 2}(1-z^2)+\left(\left(e_{\pi}\over{e_{\bar {q}}}\right)^2
-2\right)(z(1-z)+z^{2}logz)\right]\eqno (5.1)$$
and gives a contribution to the anomalous dimension of the first moment
$$\gamma_{1}^{\bar {q}q}\simeq {2\alpha_{\pi}\over 9}\left[1+{1\over 4}\left(e_{\pi}\over{e_{\bar {q}}}
\right)^2\right]\eqno(5.2)$$
This then points to a detectable evolution of the nonsinglet first moment
and, hence, of the Gottfried sum.The anomalous dimension (5.2), obtained by
retaining the large $Q^2$ approximation, is not dependent on $Q^2$ and, therefore,
cannot conform to the observed departure from the Gottfried sum rule.The $Q^2$
dependence appears when one considers the low $Q^2$ corrections.
\medskip
\leftline{\bf{6.\quad Conclusion.}}\medskip
 The simple chiral potential model considered here leads to a nontrivial
evolution of the nonsinglet moments even in the large $Q^2$ approximation.
This is  a nonperturbative phenomenon in QCD. The phenomenological model arrives at it
by allowing for emission of an isovector meson by a quark.
The detailed results for the evolution of the Gottfried sum with low $Q^2$
corrections will be reported elsewhere.
\bigskip
One of us (K.R.-M.) acknowledges financial support from the Council of Scientific
and Industrial Research during this work.
\vfill\eject
{\bf References}

\item{[1]} P.Amaudruz et al., Phys. Rev. Lett. 66,2717(1991);\par
~S.Forte, Phys.Rev. D47,1842(1993)
\item{[2]} A.Signal,A.W.Schreiber and A.W.Thomas, Mod.Phys.Lett.A6,271(1991)
\item{[3]} A.W.Thomas, Nucl.Phys.A532,171(1991)
\item{[4]} S.Kumano and J.T.Londergan, Phys.Rev.D44,717(1991)
\item{[5]} E.J.Eichten, I.Hinchliffe and C.Quigg, Phys.B425,516(1994)
\item{[6]} S.N.Jena and S.Panda, J.Phys.G:Nucl.Part.Phys.18,273(1992)
\item{[7]} S.N.Jena and S.Panda,Mod.Phys.Lett.A8,607(1993)
\item{[8]} N.Barik, B.K.Das and M.Das, Phys.Rev.D31,1652(1985)
\item{[9]} N.Barik and M.Das, Phys.Rev.D33,172(1986)
\item{[10]} R.Tegen,M.Schedle and W.Weise, Phys.Lett.125B,9(1983)
\item{[11]} R.Tegen and W.Weise,Z.Phys. A314,357(1983)
\item{[12]} S.Theberge and A.W.Thomas,Nucl.Phys.A393,252(1983)
\item{[13]} A.W.Thomas, S.Theberge, and G.A.Miller, Phys.Rev.D24,216(1981)
\item{[14]} P.Leal Ferreira,Lett.Nuovo Cimento 20,157(1977)
\item{[15]} P.Leal Ferreira and N.Zaguary, \it{ibid~}\rm  20,511(1977)
\item{[16]} G.A.Miller, in \it{International Review of Nuclear Physics}\rm
\par (Vol.1), ed. W.Weise (World Scientific, Singapore,1984)
\end